\documentclass[twocolumn,showpacs,floatfix,amsmath,amssymb,pre]{revtex4}
\usepackage{graphicx}
\usepackage{dcolumn}
\usepackage{bm}
\def\tr{t_{\rm r}}

\begin{document}
\title{Pathways of activated escape in periodically modulated systems}

\author{D. Ryvkine and M. I. Dykman}
\affiliation{Department of Physics and Astronomy, Michigan State University}
\date{\today}
\begin{abstract}
We investigate dynamics of activated escape in periodically
modulated systems. The trajectories followed in escape form
diffusion broadened tubes, which are periodically repeated in time.
We show that these tubes can be directly observed and find their
shape. Quantitatively, the tubes are characterized by the
distribution of trajectories that, after escape, pass through a
given point in phase space for a given modulation phase. This
distribution may display several peaks separated by the modulation
period. Analytical results agree with the results of simulations of
a model Brownian particle in a modulated potential.
\end{abstract}
 \maketitle

\section{Introduction}
\label{sec:Intro}

The problem of noise-activated escape in periodically modulated
systems is of interest to many areas of physics and applications,
from Josephson junctions \cite{Devoret1987,Turlot1998,Siddiqi2005}
and nano- and micromechanical systems
\cite{Aldridge2005,Stambaugh2005} to epidemics
\cite{Schwartz2004}. The theory of activated escape should answer
two closely related questions: what is the escape rate and how
does the system move during escape? Fluctuational trajectories
leading to escape or interstate switching are of significant
interest and have been extensively studied in recent years using
various numerical techniques, see Refs.
\onlinecite{Schwartz2004,Moroni2004,Zuckerman2004,E2002} and
papers cited therein. Even though escape is a random event, the
probabilities of following different paths are strongly different.
Therefore most likely the system follows a certain pathway, i.e.,
its trajectory is close to the most probable escape path (MPEP),
see Ref.~\onlinecite{Soskin2003} and references therein.

The distribution of escape trajectories can be conveniently
characterized by the prehistory probability distribution (PPD), a
quantity accessible to direct experimental measurements. It is
obtained by recording trajectories of the system that lead to escape
and superimposing the trajectories that, after the system has
escaped, pass through a small vicinity of a point $q_f$ (for a
certain modulation phase, in the case of a periodically modulated
system). The point $q_f$ is chosen in the phase space sufficiently
far behind the boundary of the basin of attraction to the initially
occupied metastable state. Formally, the PPD $p_h(q,t|q_f,t_f)$ is
the probability density for the system to have passed through a
point $q$ at an instant $t$ provided it had been fluctuating about
the metastable state for a long time and passed $q_f$ at a later
time $t_f$, $t_f>t$ \cite{Dykman1992d}.

The PPD $p_h(q,t|q_f,t_f)$ should peak at $q$ lying on the optimal
fluctuational path that leads to $q_f$. Therefore it ``maps out''
optimal paths. For stationary systems this has been directly
confirmed by extensive simulations
\cite{Luchinsky1997b,Morillo1997,Arrayas1998} and also in laser
experiments \cite{Hales2000}.  In such systems escape can occur at
any time, with equal probability, therefore the tube of paths
around the optimal escape path is broad.

\begin{figure}[tp]
\includegraphics[width=3in]{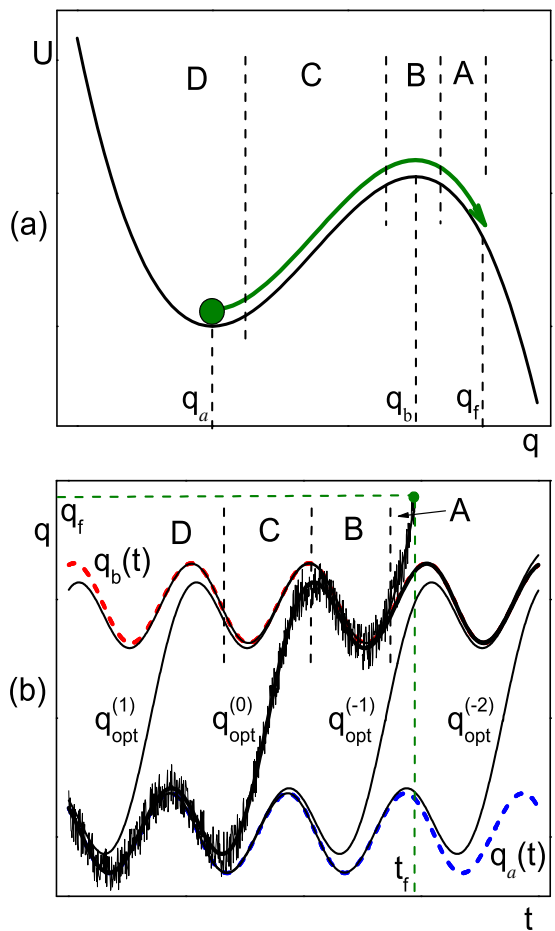}
\caption{(Color online) Activated motion leading to detection of an
escaped particle at point $q_f$ at time $t_f$ (schematically).
Panels (a) and (b) illustrate escape from a static and a
periodically modulated potential well, respectively. In the latter
case $q_a(t)$ and $q_b(t)$ are the periodic stable state and the
basin boundary. The trajectories $q_{\rm opt}^{(n)}(t)$
($n=1,\ldots,-2$) are the periodically repeated most probable escape
paths. The four major stages of motion A, B, C, and D  on the way to
$q_f$ are discussed in the text. } \label{fig:scheme}
\end{figure}

In the present paper we study escape pathways in periodically
modulated systems. Modulation synchronizes escape events. This can
be easily understood for escape from a slowly modulated potential
well. Here, escape is most likely to occur when the instantaneous
potential barrier $\Delta U(t)$ is at its lowest, once per period,
cf. Fig.~\ref{fig:scheme}(a). If the modulation amplitude of $\Delta
U(t)$ significantly exceeds the noise intensity $D$ ($D=k_BT$ for
thermal noise), escape events are synchronized very strongly. Strong
synchronization persists even where the modulation is not slow and
the adiabatic picture in which the escape rate is determined by the
instantaneous barrier height does not apply
 \cite{Smelyanskiy1997c,Lehmann2000a,Maier2001a,Dykman2005a,Ryvkine2005}.
As a result of escape synchronization, there is one MPEP per period.

\begin{figure}[tp]
\includegraphics[width=3.4in]{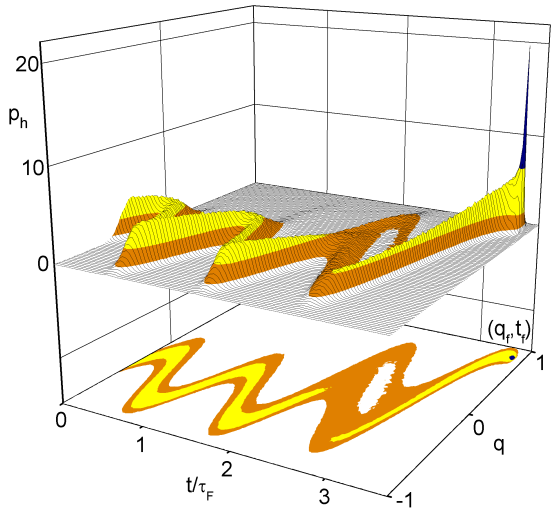}
\caption{(Color online) The prehistory probability density (PPD)
$p_h(q,t|q_f,t_f)$ and its contour plot for a noise-driven
overdamped system with equation of motion $\dot q=q^2-0.25
+A\cos\omega_Ft+ f(t)$, where $f(t)$ is white noise of intensity
$D$. The parameters are $A=0.7$, $\omega_F=2$, $D=0.01$, $q_f=0.8$,
$t_f = (\tau_F/2)({\rm mod}\,\tau_F)$, where $\tau_F=2\pi/\omega_F$
is the modulation period. The shadowing (color code on line)
corresponds to the 4 regions of the height of the distribution
separated by the values $p_h=0.5, 2, 7$.} \label{fig:3D}
\end{figure}

In turn, as we show, the prehistory probability distribution
displays a sharp narrow peak as a function of $q$ for a given $t$.
This peak lies on the MPEP. Respectively, in $(q,t)$ space the PPD
displays a narrow ridge centered at the MPEP, cf. Fig.~\ref{fig:3D}.
Moreover, the PPD may have several narrow peaks [ridges in $(q,t)$
space]. Their width gives the width of the distribution of escape
trajectories. It is determined by the typical diffusion length $l_D=
(2D\tr)^{1/2}$, where $\tr$ is the relaxation time of the system.
This is qualitatively different from the shape of the PPD in
stationary systems  \cite{Hales2000}. Of course, in modulated
systems along with the final point $q_f$ through which the system
passes one should fix the modulation phase when the passage happens.
It is given by the passage time $t_f$(mod $\tau_F$), where $\tau_F$
is the modulation period.

The occurrence of a narrow peak of the escape trajectories
distribution can be understood from the qualitative picture of
motion in escape. This picture is sketched in Fig.~\ref{fig:scheme}
for a system with one dynamical variable $q$. Escape from a static
potential well corresponds to going over the barrier top $q_b$ from
the vicinity of the potential minimum $q_a$, see
Fig.~\ref{fig:scheme}(a). Similarly, a modulated system escapes when
it goes over the periodic basin boundary $q_b(t)$ from a periodic
metastable state $q_a(t)$, see Fig.~\ref{fig:scheme}(b) (here and
below we assume that the noise correlation time is small; see
Ref.~\onlinecite{Dykman1990} for a more general case). Let us
suppose that the escaped particle is found at time $t_f$ at a point
$q_f$ sufficiently far behind $q_b(t)$. A typical trajectory to this
point displays four distinct sections with different types of motion
shown schematically by letters A through D.

We discuss the motion backward in time from $t_f$. Immediately
adjacent to $q_f$ is section A of the trajectory in
Fig.~\ref{fig:scheme}. Here, for small noise intensity the system
moves close to the noise-free trajectory from the vicinity of
$q_b(t)$ to $q_f$. In the case of a static potential this
corresponds to sliding down the potential slope from $q_b$ to $q_f$.
Section B corresponds to diffusion in a region of width $\sim l_D$
around $q_b(t)$. In this region noise-free motion with respect to
$q_b(t)$ is slow (in the case of a static potential the potential is
locally flat at $q_b$). Section C corresponds to motion from the
attractor to the basin boundary. The respective trajectory is close
to the most probable escape path. This motion is a result of the
large fluctuation that has led to escape detected at $q_f$. For
strongly synchronized escape, there is one MPEP per modulation
period, as shown in Fig.~\ref{fig:scheme}(b). The MPEP approaches
$q_b(t)$ asymptotically as $t\to \infty$. If the system was observed
behind $q_b(t_f)$ at time $t_f$, it has most likely arrived to the
vicinity of $q_b(t)$ along the MPEP that approached $q_b(t)$ before
$t_f$, but not too much in advance, as shown in
Fig.~\ref{fig:scheme}(b). Finally in the region D well before $t_f$
the system was fluctuating about the attractor.

The above picture suggests that the PPD will peak on the trajectory
singled out in Fig.~\ref{fig:scheme}. It also explains why the PPD
peak along $q$ axis should be narrow: in contrast to stationary
systems, which can arrive to the vicinity of $q_b$ at any time,
periodically modulated systems approach $q_b(t)$ only once per
period, and the tube of escape trajectories is narrow. We note that,
by changing $q_f, t_f$, we can switch between neighboring MPEP's,
and therefore there is a possibility for the PPD to have two and
potentially even more peaks inside the attraction basin.

As mentioned above, we are interested in the regime where escape
events are strongly synchronized by the modulation. It means that
the probability density to find the escaped system at a point $q_f$
behind the basin boundary, cf. Fig.~\ref{fig:scheme}, displays sharp
peaks as a function of the observation time $t_f$. These peaks are
periodically repeated in time. They were studied in
Ref.~\onlinecite{Ryvkine2005}. We will analyze the PPD for the time
$t_f$ close to the maximum of the escape probability peak. This
choice is justified, because the corresponding PPD characterizes the
most probable escape trajectories. Otherwise the PPD would be formed
by trajectories with exponentially smaller probabilities that are
very rarely followed in escape.

Interestingly, the condition that the tubes of escape trajectories
be narrow in time and space requires that the modulation frequency
$\omega_F=2\pi/\tau_F$ lie within a range limited both from below
and from above. To understand the lower limit we note that, as
mentioned above, for slow modulation escape occurs every period
around the time $t_{\rm m} + k\tau_F$ where the height of the
instantaneous potential barrier $\Delta U(t)$ is at its minimum
($k=0,\pm 1,\ldots$). The typical width of the time window for
escape $\Delta t$ is given by the condition $|\Delta U(t_{\rm
m}\pm\Delta t) - \Delta U(t_{\rm m})|\lesssim D$, which leads to
$\Delta t=[D/\Delta\ddot U(t_{\rm
m})]^{1/2}\sim\omega_F^{-1}[D/\Delta U(t_{\rm m})]^{1/2}$. If
$\Delta t$ exceeds the typical duration of motion in escape
$\tr\ln[\Delta U(t_{\rm m})/D]$, the PPD has the same shape as if
the system were escaping out of a stationary potential well of
height $\Delta U(t_{\rm m})$. In this case inside the attraction
basin the ridge of the PPD in $(q,t)$ space is broad and asymmetric,
its width along $q$ axis is independent of the noise intensity
\cite{Hales2000}.

In the opposite limit of high-frequency modulation, $\omega_F\tr \gg
1$, escape of an overdamped system is not synchronized. The dynamics
is characterized by the coordinates averaged over modulation period.
The PPD as a function of such coordinates is described by the theory
for stationary systems \cite{Hales2000}, and the PPD peak inside the
attraction basin is broad.

In the intermediate range of frequencies not only are the PPD peaks
narrow but, as mentioned above, the PPD as a function of $q$ for
given $t$ (and also of $t$ for given $q$) may display several peaks
inside the attraction basin. This happens because the motion of the
system near the basin boundary is slow. Therefore if the system is
observed behind the boundary $q_b(t)$ at a given time $t_f$, it
could have arrived to the boundary along one of a few periodically
repeated optimal escape paths, fluctuated about $q_b(t)$ for some
time, and then made a transition to $(q_f,t_f)$ over time $\sim
\tr$.

The multiple-peak structure of the PPD is a specific feature of
periodically modulated systems far from equilibrium, i.e. away from
both the adiabatic limit of slow modulation and the limit of fast
modulation. It is illustrated in Fig.~\ref{fig:3D} for a model
system. Two ridges of the PPD inside the attraction basin are
clearly resolved in this figure. Their shape is well described by
the asymptotic theory developed in this paper.

In Sec.~II below we discuss the dynamics of a periodically modulated
system and the equations for the most probable escape path. In
Sec.~III we obtain a general expression for the PPD near the basin
boundary and relate it to the distribution that describes the
quasiperiodic current away from the attraction basin, which gives
the escape rate. In Sec.~IV it is shown that the PPD is simplified
in the adiabatic limit where the modulation period is small compared
to the relaxation time. Two types of behavior may be displayed near
$q_b(t)$ in this case, depending on the ratio of the two small
parameters that characterize the dynamics. The analysis is extended
to the case of nonadiabatic driving in Sec.~V. In Sec.~VI we obtain
the central result of the paper, the PPD inside the attraction
basin. We show that it has the shape of a sum of diffusion-broadened
Gaussian peaks centered at the periodically repeated most probable
escape paths. In Sec.~VII we describe the results of simulations of
a model system and compare them to the analytical results. Sec.~VIII
provides a brief summary of the results.

\section{Escape of a periodically modulated system}
\label{sec:model}

We investigate the prehistory probability distribution for an
overdamped system characterized by one dynamical variable $q$. The
system is in a periodically modulated potential
$U(q,t)=U(q,t+\tau_F)$. Fluctuations are induced by an external
noise $f(t)$. The motion is described by the Langevin equation
\begin{equation}
\label{eq:Langevin}
\dot q = K(q,t) + f(t).
\end{equation}
\noindent Here, $K(q,t)\equiv\partial_qU(q,t)$ is the regular
periodic force. We consider the simplest case where $f(t)$ is
zero-mean white Gaussian noise with correlator $\langle
f(t)f(t')\rangle = 2D\delta (t-t')$.

In the absence of noise the system (\ref{eq:Langevin}) has a periodic
metastable state (attractor) $q_a(t)=q_a(t+\tau_F)$ and a periodic
boundary $q_b(t)=q_b(t+\tau_F)$ of the basin of attraction to
$q_a(t)$, see Fig.~\ref{fig:scheme}(b). The states $q_a(t)$ and
$q_b(t)$ are, respectively, the stable and unstable periodic solutions
of the equation $\dot{q}=K(q,t)$. For concreteness we assume that
$q_a(t) < q_b(t)$.

The PPD $p_h(q,t|q_f,t_f)$ as defined in the Introduction is the
conditional probability density (with respect to $q$) of passing
through a point $(q,t)$ in the coordinate-time space on the way from
the attractor to a point $(q_f,t_f)$. It has the form
\cite{Dykman1992d}
\begin{eqnarray}
\label{eq:PPD_def}
p_h(q,t|q_f,t_f)=\frac{\rho(q_f,t_f|q,t)\rho(q,t|q_{\rm in},t_{\rm in})}
{\rho(q_f,t_f|q_{\rm in},t_{\rm in})},
\end{eqnarray}
where $\rho(q_1,t_1|q_2,t_2)$ is the probability density (with respect
to $q_1$) of a transition from $(q_2,t_2)$ to $(q_1,t_1)$, with
$t_1>t_2$.

We assume that the noise intensity $D$ is small. Then the
period-average escape rate is $\overline W \propto \exp(-R/D)$,
where $R$ is the activation energy of escape, $R\gg D$
\cite{Smelyanskiy1997c,Lehmann2000a,Maier2001a,Dykman2005a,Ryvkine2005}.
The condition on $D$ is that ${\overline W}^{-1}$ largely exceeds
the relaxation time of the system $\tr$ and the modulation period
$\tau_F$. Eq.~(\ref{eq:PPD_def}) gives the distribution of escape
paths in a broad time range
\begin{equation}
\label{eq:condition}
\overline{W}^{-1} \gg t_f-t_{\rm in} > t-t_{\rm in} \gg t_{\rm r}, \tau_F,
\end{equation}
where the population of the attraction basin practically does not
change. For $t-t_{\rm in}\gg \tr$ the initial state $q_{\rm in}$,
which is close to the attractor $q_a(t_{\rm in})$, gets forgotten.
Then the right-hand side of Eq.~(\ref{eq:PPD_def}) becomes
independent of $t_{\rm in}$.

The functions $\rho(q,t|q_{\rm in},t_{\rm in})=\rho(q,t)$ and
$\rho(q_f,t_f|q_{\rm in},t_{\rm in})=\rho(q_f,t_f)$ give the
time-periodic probability density to find the system in states $q$ and
$q_f$, respectively. To study escape pathways we calculate the PPD for
$q_f$ outside the attraction basin, $q_f> q_b(t_f)$.

The transition probability density $\rho(q,t|q',t')$ is a solution of
the Fokker-Planck equation (FPE)
\begin{equation}
\label{eq:FPE}
\partial_t\rho=-\partial_q[K(q,t)\rho]+D\partial^2_q\rho,
\end{equation}
with the initial condition $\rho(q,t'|q',t')=\delta(q-q')$. Even for a
1D system, this equation does not have a known explicit solution
except for the case of $K$ linear in $q$. To analyze the PPD we will
have to find approximate solutions in different regions and match
them.

In the following subsections we discuss the dynamics of the system prior and
during escape.

\subsection{Dynamics near the periodic states}

We start with the dynamics close to the periodic states
$q_i(t)$, $i=a,b$. In the absence of noise it is described by the
linear equations $\delta\dot{q} = \mu_i(t)\delta q$ with
$\delta q=q-q_i(t)$ and
\begin{equation}
\label{eq:mu}
\mu_i(t)=\mu_i(t+\tau_F)=\left[\partial_qK\right]_{q_i(t)}, \qquad i=a,b.
\end{equation}
Time evolution of the deviation $\delta q$ from $q_i(t)$ is given by
\begin{eqnarray}
\label{eq:kappa}
&&\delta q(t) = \delta q(t')\kappa_i(t,t'), \\
&&\kappa_i(t,t')=\exp\left[ \int_{t'}^td\tau\mu_i(\tau) \right],
\qquad i=a,b.\nonumber
\end{eqnarray}
Because of the periodicity of $\mu_i(t)$, the Floquet multipliers
\begin{equation}
\label{eq:Floquet}
M_i=\kappa_i(t+\tau_F,t)=\exp\left(\bar\mu_i\tau_F\right), \qquad i=a,b,
\end{equation}
are independent of time, with $M_a<1$ and $M_b>1$. In
Eq.~(\ref{eq:Floquet}) $\bar{\mu}_i$ $(i=a,b)$ is the period-average
value of $\mu_i(t)$, with $\bar{\mu}_a<0$, $\bar{\mu}_b>0$. The
relaxation time of the system can be chosen as $t_{\rm
r}=\bar{\mu}_b^{-1}\sim|\bar{\mu}_a|^{-1}$.

Weak noise leads to fluctuations about the attractor, which have
Gaussian distribution near the maximum in the time range
(\ref{eq:condition}). From Eq.~(\ref{eq:FPE}) this distribution has
the form
\begin{equation}
\label{eq:PPD_limit}
\rho_a(q,t)=G\left(q-q_a(t);\sigma_a(t)\right),
\end{equation}
with
\begin{equation}
\label{eq:Gaussian_peak}
G(x;\sigma) = (2\pi D \sigma^2)^{-1/2}\exp(-x^2/2D\sigma^2).
\end{equation}
The variance $\sigma_a^2(t)$ is periodic in time. It is given by the
expression
\begin{equation}
\label{eq:sigma_a}
\sigma^2_i(t)=2\left|M_i^{-2}-1\right|^{-1}\int\nolimits_0^{\tau_F}dt_1 \kappa_i^{-2}(t+t_1,t)
\end{equation}
with $i=a$.

\subsection{Most probable escape paths}

Along with small fluctuations, there also happen occasional large
deviations from $q_a(t)$, including escape from the attraction
basin. In a broad parameter range escape events are very strongly
synchronized with the modulation, see Ref.~\onlinecite{Ryvkine2005}
and papers cited therein. Because large fluctuations have small
probabilities, in escape the system closely follows the trajectory
that is least improbable among all possible escape trajectories. As
mentioned in the Introduction, it is usually called the most
probable escape path, $q_{\rm opt}(t)$. Inside the attraction basin,
escape trajectories lie within periodically repeated narrow tubes.
The tubes have a width $\sim l_D$ and are centered at the MPEP's
$q_{\rm opt}^{(n)}(t)\equiv q_{\rm opt}(t+n\tau_F)$, $n=0,\pm
1,\ldots$.

The MPEP's provide a solution to the variational problem of
maximizing the probability of a fluctuation in which the system
moves from $q_a(t)$ to $q_b(t)$. This problem can be mapped onto the
problem of dynamics of an auxiliary Hamiltonian system. The latter
is described by the Wentzell-Freidlin Hamiltonian $H(p,q;t) =
p^2+pK(q,t)$ \cite{Freidlin_book}. Its equations of motion have the
form
\begin{equation}
\label{eq:aux_system}
\dot{q} = K(q,t) + 2p, \qquad \dot{p} = -p\,\partial_qK.
\end{equation}
The MPEP's correspond to the minimal action heteroclinic Hamiltonian
trajectories $(q_{\rm opt}(t),p_{\rm opt}(t))$
\cite{Graham1984a,Smelyanskiy1997c}. They start for $t\to -\infty$ at
the periodic hyperbolic state $(q_a(t),p=0)$ of the auxiliary system
and for $t\to \infty$ approach its another periodic hyperbolic state,
$(q_b(t),p=0)$.

Well outside the diffusion regions around the periodic states
$q_{a,b}(t)$ the motion along the MPEP is fast. It is seen from
Eq.~(\ref{eq:aux_system}) that the system moves between these
regions over time $\tr$. Close to $q_b(t)$ the system is slowed
down, and in the region $|q-q_b(t)|\lesssim l_D$ the motion is
dominated by diffusion. The duration of staying in the vicinity of
$q_b(t)$ can be obtained by linearizing the equation of motion
(\ref{eq:Langevin}) near $q_b$ and is given by the Suzuki time
$t_S\sim \bar{\mu}_b^{-1}\ln(\Delta q/l_D)$
\cite{Suzuki1977a,Suzuki1977}, where $\Delta q =
\min_t|q_b(t)-q_a(t)|$ is the typical distance between the periodic
states.

Periodically modulated systems are advantageous as they allow one to
observe, via the prehistory distribution, both the fast motion along
the MPEP and the slow motion near the unstable state. As we show
below, the peak of the PPD does not display broadening due to
diffusion near $q_b$, as does the PPD in the absence of modulation
\cite{Hales2000}.

\section{Prehistory Probability Distribution near the basin boundary}
\label{sec:boundary}

We will calculate the PPD in the regime of strong synchronization of
escape \cite{Dykman2005a,Ryvkine2005}. In this regime the
probability distribution $\rho(q,t)$ of finding a particle behind
the basin boundary has the form of sharp periodic pulses as a
function of $t$. It is most interesting to find the PPD for a final
point $(q_f,t_f)$ on the $(q,t)$ plane near the center of such a
pulse.

We will assume that the point $(q_f,t_f)$ is sufficiently far from
the diffusion-dominated layer (region B in Fig. \ref{fig:scheme})
around the basin boundary, so that the distance to the boundary is
$Q_f= q_f-q_b(t_f)\gg l_D$. This condition is usually realized in
experiments, where the position of a particle detector is chosen so
as to ensure that the detected particles have practically no chance
to return to the attraction basin.  At the same time we assume for
convenience that the final point is still in the harmonic region,
$Q_f\ll \Delta q$, in which case the motion between $q_b$ and $q_f$
can be described by linearizing equation (\ref{eq:Langevin}) in
$Q=q-q_b(t)$.

We will start the analysis of the PPD $p_h(q,t|q_f,t_f)$ with the
case where not only $q_f$ but also the point $(q,t)$ through which
the trajectory of interest has passed is also in the harmonic region
around the basin boundary $q_b(t)$.

\subsection{Transition probability density}

As seen from Eq.~(\ref{eq:PPD_def}), finding the PPD requires
calculating the transition probability density $\rho(q_f,t_f|q,t)$.
The Fokker-Planck equation for $\rho(q_f,t_f|q,t)$ can be linearized
near $q_b(t)$,
\begin{equation}
\label{eq:FPE_linearized}
\partial_{t_f}\rho=-\mu_b(t_f)\partial_{Q_f}(Q_f\rho)+D\partial^2_{Q_f}\rho,
\end{equation}
where $\rho = \rho(q_f,t_f|q,t)$ and $Q_f=q_f-q_b(t_f)$.

The solution of Eq.~(\ref{eq:FPE_linearized}) can be sought in the
form of a Gaussian distribution
\begin{equation}
\label{eq:FPE_linearized_sol}
\rho=G\left(Q_f-\tilde{Q}(t_f);\tilde{\sigma}(t_f)\right),
\end{equation}
with $G(x;\sigma)$ given by Eq.~(\ref{eq:Gaussian_peak}).
Eq.~(\ref{eq:FPE_linearized}) will be satisfied provided the
hitherto unknown functions $\tilde\sigma, \tilde Q$ obey the
equations
\begin{equation}
\label{eq:Gaussian_coeff}
\frac{d{\tilde{\sigma}}^2}{dt_f}=2\mu_b(t_f)\tilde{\sigma}^2+2,
\quad \frac{d\tilde{Q}}{dt_f}=\mu_b(t_f)\tilde{Q}.
\end{equation}
The initial conditions for these equations follow from the condition
$\rho(q_f,t|q,t)=\delta(q_f-q)$. They have the form
$\tilde{\sigma}^2(t)=0$ and $\tilde{Q}(t)= Q =q-q_b(t)$. Then the
solution of Eqs.~(\ref{eq:Gaussian_coeff}) is
\begin{eqnarray}
\label{eq:Gaussian_coeff_sol}
&&\tilde{\sigma}^2(t_f)=2\int\nolimits_{t}^{t_f}d\tau\kappa_b^2(t_f,\tau),
\nonumber\\
&&\tilde{Q}(t_f)=Q\kappa_b(t_f,t).
\end{eqnarray}
Finally, using the function
\begin{equation}
\label{eq:Sigmab_def}
\sigma_f^2(t_f,t)=\kappa_b^{-2}(t_f,t)\tilde{\sigma}^2(t_f)=
2\int\nolimits_{t}^{t_f}d\tau\kappa_b^{-2}(\tau,t),
\end{equation}
we can write the distribution in the form
\begin{eqnarray}
\label{eq:rho_harmonic}
\rho=\kappa_b^{-1}(t_f,t)G\left(Q-Q_f\kappa_b^{-1}(t_f,t); \sigma_f(t_f,t)\right).
\end{eqnarray}

The function $Q_f\kappa_b^{-1}(t_f,t)$ has a simple meaning.  Consider
the noise-free trajectory that passes through the point $Q_f$ at time
$t_f$. This trajectory should have passed through the point
$Q_f\kappa_b^{-1}(t_f,t)$ at time $t$. As expected, the transition
probability (\ref{eq:rho_harmonic}) is maximal for $Q$ coinciding with
this point.

The function $\sigma_f^2$, Eq.~(\ref{eq:Sigmab_def}), is simply
related to the  function $\sigma_b^2(t)$, Eq.~(\ref{eq:sigma_a}),
introduced earlier,
\begin{equation}
\label{eq:sigma_f_and_sigma_b}
\sigma_b^2(t) - \kappa_b^{-2}(t_f,t)\sigma_b^2(t_f)=\sigma_f^2(t_f,t).
\end{equation}
We will use this relation in what follows.

\subsection{General expression for the PPD near $q_b(t)$}

From Eq.~(\ref{eq:PPD_def}), the PPD is determined by the product of
the transition probability (\ref{eq:rho_harmonic}) and the ratio of
the quasiperiodic distributions $\rho(q,t)/\rho(q_f,t_f)$. The
distribution $\rho(q,t)$ close to $q_b(t)$ was found earlier
\cite{Dykman2005a,Ryvkine2005}. It has the form
\begin{eqnarray}
\label{eq:rho_Laplace}
&&\rho(q,t) = \int_0^{\infty}dp\, \rho'(p,Q,t),\qquad Q=q-q_b(t),\nonumber\\
&&\rho'(p,Q,t)=
\frac{\mathcal{E}}{\sqrt{D}}\exp\!
\left\{-\frac{1}{D}\left[\frac{p^2\sigma_b^2(t)}{2}
+pQ + s(\phi)\right]\right\},\nonumber\\
&&\phi(p,t) = \Omega_F\ln\left[\frac{p\kappa_b(t,t')}{\bar{\mu}_bl_D}\right],
\qquad \Omega_F=\omega_F/\bar{\mu}_b.
\end{eqnarray}
Here $\mathcal{E}$ and $t'$ are constants, $s(\phi)=s(\phi+2\pi)$ is
a zero-mean $2\pi$-periodic function, and $\Omega_F$ is the
dimensionless modulation frequency.

The function $s(\phi)$ in Eq.~(\ref{eq:rho_Laplace}) plays the role
of an instantaneous modulation-induced change of the activation
energy. In the regime  of strong synchronization of escape the
minimal value $s_{\rm m}$ of $s(\phi)$ satisfies the condition
$|s_{\rm m}|\gg D$. The minima of $s(\phi)$ lie on the optimal
escape paths \cite{Dykman2005a,Ryvkine2005}, $p=p_{\rm
opt}^{(n)}(t)\equiv p_{\rm opt}(t+n\tau_F)$. Here $n=0,\pm 1,\ldots$
enumerates periodically repeated MPEP's, see
Fig.~\ref{fig:scheme}(b); we set $(q_{\rm opt}^{(0)}(t),p_{\rm
opt}^{(0)}(t))=(q_{\rm opt}(t),p_{\rm opt}(t))$.

Near the basin boundary $q_b(t)$ the optimal paths satisfy the
linearized equations (\ref{eq:aux_system}) and evolve in time as
\begin{eqnarray}
\label{eq:optimal_near_b}
&&p_{\rm opt}(t)=\kappa_b^{-1}(t,t')p_{\rm opt}(t'),
\quad Q_{\rm opt}(t)=-\sigma_b^2(t)p_{\rm opt}(t), \nonumber\\
&&Q_{\rm opt}(t) = q_{\rm opt}(t) - q_b(t),
\quad |Q_{\rm opt}|\ll \Delta q.
\end{eqnarray}
Eq.~(\ref{eq:optimal_near_b}) describes how a given optimal path
approaches $(q_b(t),p=0)$ for $t\to \infty$. The parameter $t'$ is
determined by matching to one of the periodically repeated
trajectories (\ref{eq:aux_system}) that start from $(q_a(t),p=0)$ for
$t\to -\infty$.

Expanding the function $s(\phi)$ around its minima at the
periodically repeated MPEP's, to second order in
$\phi(p,t)-\phi\bigl(p_{\rm opt}^{(n)}(t),t\bigr)$ we obtain the
probability distribution (\ref{eq:rho_Laplace}) as a sum of
contributions from the MPEP's,
\begin{eqnarray}
\label{eq:rho_Laplace_expand}
&&\rho'(p,Q,t) = \frac{\mathcal{E}e^{-s_{\rm m}/D}}{\sqrt{D}}
\sum_n\exp\left[-r^{(n)}(p,Q,t)/D\right], \nonumber \\
&&r^{(n)}=pQ+\frac{p^2\sigma_b^2(t)}{2}+
\frac{\Omega^2_Fs''_{\rm m}}{2}\ln^2\left[\frac{p}{p_{\rm opt}^{(n)}(t)}\right].
\end{eqnarray}
Here $s''_{\rm m}\equiv(d^2s/d\phi^2)_{\rm m}$ is the curvature of the
function $s$ at the minimum. For strong synchronization, where
$|s_{\rm m}|\gg D$, we have $s''_{\rm m}\gg D$ as well, which is a
consequence of $s(\phi)$ being a zero-mean periodic function. The
quantity $s''_{\rm m}$ can be found \cite{Dykman2005a,Ryvkine2005}, along
with the constant $\mathcal{E}\exp(-s_{\rm m}/D)$, by matching the
periodic distribution (\ref{eq:rho_Laplace}) to the distribution well
inside the attraction basin.

It follows from Eq.~(\ref{eq:optimal_near_b}) that near the basin boundary
\begin{equation}
\label{eq:p_near_b}
p^{(n+k)}_{\rm opt}(t) = p^{(n)}_{\rm opt}(t+k\tau_F) =
M_b^{-k}p^{(n)}_{\rm opt}(t).
\end{equation}
With account taken of the relation $M_b=\exp(2\pi/\Omega_F)$, this
leads to the expression
\begin{equation}
\label{eq:r_near_b}
r^{(n+k)} = r^{(n)} + 2\pi k\Omega_Fs''_{\rm
m}\ln\left[\frac{p}{p^{(n)}_{\rm opt}(t)} \right] + 2\pi^2k^2s''_{\rm
m}.
\end{equation}

Combining Eqs.~(\ref{eq:rho_harmonic}) and
(\ref{eq:rho_Laplace_expand}) and using
Eq.~(\ref{eq:sigma_f_and_sigma_b}) we obtain for the PPD near the
basin boundary,
%
\begin{eqnarray}
\label{eq:PPD_b_general}
&&p_h(q,t|q_f,t_f) \nonumber\\
&&=\kappa_b^{-1}(t_f,t)
G\left(Q-Q_f\kappa_b^{-1}(t_f,t);\sigma_f(t_f,t)\right)\nonumber\\
&&\times\frac{ \sum_n\int_0^{\infty}dp\exp\left[-r^{(n)}(p,Q,t)/D\right] }
{\sum_n\int_0^{\infty}dp\exp\left[-r^{(n)}(p,Q_f,t_f)/D\right]}.
\end{eqnarray}
This expression gives the general form of the PPD near the basin
boundary. We note that, even though the equations of motion near
$q_b(t)$ can be linearized,  the PPD is generally non-Gaussian. This
distortion is an important feature of escape dynamics.

\section{Adiabatic regime near the basin boundary}
\label{sec:adiabatic}

We start the analysis with the case of slow modulation, $\Omega_F\ll
1$, where the motion can be described in the adiabatic
approximation. As we show below, in the adiabatic regime and for
strong synchronization, $s''_{\rm m}\gg D$, only one term
contributes to each of the sums in Eq.~(\ref{eq:PPD_b_general}).
The shape of the PPD in this case is determined by the parameter
\begin{equation}
\label{eq:theta}
\theta=\Omega_F^2s''_{\rm m}/D.
\end{equation}

We call $\theta$ the distortion parameter. This is because for
$\theta\ll 1$ the escape current has a form of Gaussian peaks,
whereas for larger $\theta$ the peaks of the current become
non-Gaussian \cite{Dykman2005a,Ryvkine2005}. Formally, $\theta$ is
equal to the ratio of two small parameters, the squared reduced
modulation frequency $\omega_F/\bar\mu_b = \Omega_F$ and the noise
intensity scaled by the effective modulation strength $D/s''_{\rm
m}$.

The physical meaning of $\theta$ can be understood in the
following way. In the adiabatic picture one usually thinks of
escape as occurring in the instantaneous potential $U(q,t)$. Most
likely it happens once per period at the time $t_{\rm m}$ when the
barrier height $\Delta U(t)=U(q_b(t),t)-U(q_a(t),t)$ is minimal,
cf. Fig.~\ref{fig:scheme}. As explained in the Introduction, the
typical width of the time window for escape $\Delta t$ is
determined by the condition $[d^2\Delta U/dt^2]_{\rm m}(\Delta
t)^2=D$, where the subscript $m$ indicates that the derivative is
evaluated for $t=t_{\rm m}$.  The parameter $s''_{\rm m}$ is (see
Ref.\onlinecite{Ryvkine2005})
\[s''_{\rm m}=[d^2\Delta U/dt^2]_{\rm m}/\Omega_F^2\mu_b(t_{\rm m})\sim
[d^2\Delta U/dt^2]_{\rm m}/\omega_F^2.\]
Therefore the parameter $\theta\sim [(\Delta t)^2\bar\mu_b^2]^{-1}$
is the squared ratio of the relaxation time of the system
$\tr=\bar\mu_b^{-1}$ to $\Delta t$. It shows whether the system
moves fast enough to escape while the barrier remains at its minimum
or the barrier noticeably changes during escape leading to a delay
of escape with respect to $t_{\rm m}$ and a distortion of the tube
of escape trajectories.

Because the PPD evolves over time $\sim \tr$, of interest is the
time range $|t-t_f|\lesssim\tr\ll \tau_F$. Moreover, both $t$ and
$t_f$ should be close to $t_{\rm m}$. Then the instantaneous
relaxation rate $\mu_b(t)$ can be approximated by its value
$\mu_{b{\rm m}}\equiv\mu_b(t_{\rm m})$.  The functions $\kappa_b$
and $\sigma_f^2$ become
\begin{eqnarray}
\label{eq:ad_functions}
&&\kappa_{b{\rm m}}(t_f,t) = \exp\left[\mu_{b{\rm m}}(t_f-t)\right],
\qquad \nonumber \\
&&\sigma_{f{\rm m}}^2(t_f,t) = \sigma_{b{\rm m}}^{2}
\left\{1-\exp\left[-2\mu_{b{\rm m}}(t_f-t)\right]\right\},
\end{eqnarray}
where
\[\sigma_b^2(t)\approx\sigma_b^2(t_{\rm m})\equiv
\sigma_{b{\rm m}}^2=1/\mu_{b{\rm m}}.\]

\subsection{Weak distortion, $\theta \ll 1$}
\label{subsec:weak_distortion}

We consider first the limit of weak distortion, $\theta\ll 1$ (more
precisely, the weak distortion condition has the form
$\theta\ln^2(s''_{\rm m}/D)\ll 1$, see below). In this limit one can
think of motion in the fully adiabatic way, assuming that it occurs
in a quasistatic potential $U(q,t_{\rm m})$. The periodic
distribution behind the attraction basin $\rho(q_f,t)$, which is
proportional to the instantaneous escape rate, has a form of
periodic in time Gaussian pulses, with width $\sim (D/s_{\rm
m}'')^{1/2}\tau_F$ \cite{Ryvkine2005}.

For small $\theta$ expression (\ref{eq:PPD_b_general}) for the PPD
can be simplified. We will start with the analysis of the
denominator in this expression. If there were no term
$\propto\Omega_F^2s''_{\rm m}/D=\theta$ in $r^{(n)}(p,Q_f,t_f)/D$,
the typical values of $p$ contributing to the integral over $p$
would be $\sim D/Q_f$. For such $p$ there may be only one $n$ for
which the term $\theta\ln^2[p/p_{\rm opt}^{(n)}(t_f)]$ in
$r^{(n)}/D$ is small, whereas for all other $n$ it is $\propto
s_{\rm m}''/D$ [cf. Eq.~(\ref{eq:r_near_b})], making the integrand
exponentially small. A similar argument applies to the numerator
in Eq.~(\ref{eq:PPD_b_general}), except that, depending on $Q$,
the typical values of $p$ are of order of $(D/\mu_{b{\rm
m}})^{1/2}, D/Q$, or $-\mu_{b{\rm m}}Q$.

Keeping only the leading term in the sums in the numerator and
denominator of Eq.~(\ref{eq:PPD_b_general}) and disregarding
corrections $\propto\theta$, we obtain
\begin{eqnarray}
\label{eq:PPD_b_small_theta}
&&p_h(q,t|q_f,t_f) =
G\left(Q-Q_f\kappa_{b{\rm m}}^{-1}(t_f,t);\sigma_{f{\rm m}}(t_f,t)\right)
\nonumber\\
&&\times\frac{Q_f\kappa_{b{\rm m}}^{-1}(t_f,t)}
{\sqrt{2D\sigma_{b{\rm m}}^2}}
\exp\left[\frac{Q^2}{2D\sigma_{b{\rm m}}^2}\right]{\rm
erfc}\left[\frac{Q}{\sqrt{2D\sigma_{b{\rm m}}^2}}\right].
\end{eqnarray}
In deriving this expression we also took into account that the final
point $q_f$ is sufficiently far from the diffusion-dominated region
behind the basin boundary, $Q_f\gg l_D$, and disregarded corrections
$\sim l_D^2/Q_f^2$. We note that the terms $\propto \theta$ in
$r^{(n)}$ in the numerator and denominator in
Eq.~(\ref{eq:PPD_b_general}) have a logarithmic factor. This factor
may become large in the weak-noise limit. One can show that for the
corrections $\propto \theta$ to be small it is necessary that
$\theta\ln^2(s''_{\rm m}/D)\ll 1$. This condition is equivalent to
the inequality $\Delta t\gg t_S$, where $\Delta t$ is the
characteristic time window within which the barrier height is
practically constant and $t_S$ is the Suzuki time, see
Eq.~(\ref{eq:Suzuki_time}) below.

Equation~(\ref{eq:PPD_b_small_theta}) for the PPD is further
simplified for such times that the point $Q_f\kappa_b^{-1}(t_f,t)$
is far behind the basin boundary compared to $l_D$. This is the
point that the noise-free trajectory arriving at $Q_f$ at time $t_f$
passes at time $t$. For such $t$, the PPD as a function of $Q$ has
the form of a Gaussian peak with variance $D\sigma_f^2(t_f,t)$. This
means that the trajectories arriving to the point $Q_f$ are close to
the noise-free trajectory. The tube of these trajectories is
diffusion-broadened, with width $\propto [D(t_f-t)]^{1/2}$ for small
$t_f-t$.

Because for $\theta\ll 1$ escape as a whole occurs in the quasistatic
potential $U(q,t_{\rm m})$, the full calculation of the PPD described
in Section~VI leads to the same result as in the case of escape
in a stationary potential studied earlier~\cite{Hales2000}.  With
increasing $t_f-t$ the peak of the PPD (\ref{eq:PPD_b_small_theta})
crosses the basin boundary and enters the attraction basin. This is
the slowest part of the PPD evolution.  Its duration is determined by
the Suzuki time
\begin{equation}
\label{eq:Suzuki_time}
t_S=\tr\ln(s_{\rm m}''/D).
\end{equation}
The peak is sharply broadened in this region. Deep on the intrawell
side, $-Q\gg l_D$, the width of the peak becomes independent of $D$,
i.e. parametrically larger than the diffusion-limited width $\sim
l_D$.  As $t_f-t$ increases further, the PPD peak approaches the
attractor and narrows down, with the width becoming again
diffusion-limited. This part of the evolution takes $\sim\tr$.

\subsection{Strong distortion, $\theta \gg 1$}

The shape of the PPD near the maximum changes dramatically in the
range where the modulation is still dynamically slow, $\Omega_F\ll
1$, but the distortion parameter $\theta$ becomes large, $\theta\gg
1$. Here, the shape of the potential barrier for escape changes as
the particle crosses the diffusion region around the basin boundary.
This leads to a strong change of the PPD compared to the picture
based on the quasistatic barrier that was discussed before.

For $\theta\gg 1$ in an important range of $Q,t$ and $Q_f,t_f$ the
integrands in the numerator and denominator in
Eq.~(\ref{eq:PPD_b_general}) have sharp extrema as functions of $p$
for $p=p^{(n)}_{\rm opt}$  (a more precise condition is specified
below). Integration over $p$ can be done by the steepest descent
method. It gives the PPD in the form of a Gaussian distribution over
$Q$ with time-dependent center $Q_{n_0}$ and variance $D\sigma_f^2$,
\begin{eqnarray}
\label{eq:PPD_b_large_theta_ad} &&p_h(q,t|q_f,t_f) =
G\left(Q-Q_{n_0}; \sigma_f(t_f,t)\right),
\end{eqnarray}
where
\begin{eqnarray}
\label{eq:Q_n_0}
Q_n&\equiv& Q_{n}(t|q_f,t_f)\nonumber\\
&=& Q^{(n)}_{\rm opt}(t)+\left(Q_f- Q^{(n)}_{\rm
opt}(t_f)\right)\kappa_b^{-1}(t_f,t)
\end{eqnarray}
and $n_0\equiv n_0(q_f,t_f)$. Here we have taken into account that
the major contribution to each of the sums in
Eq.~(\ref{eq:PPD_b_general}) comes from one term, with $n=n_0$.

It is clear from the analysis of the denominator in
Eq.~(\ref{eq:PPD_b_general}) that the value of $n_0$ is determined
by the condition that $p^{(n_0)}_{\rm opt}(t_f)Q_f/D\sim 1$ for the
$n_0$th optimal path. The very existence of such $n_0$ follows from
the fact that we consider $t_f$ for which the probability density
$\rho(q_f,t_f)$ is close to its maximum over $t_f$; this maximum is
reached once per period for $p_{\rm opt}^{(n)}(t_f)Q_f/D=1$
\cite{Dykman2005a,Ryvkine2005}. The terms with $|n-n_0|\geq 1$ in
the denominator in Eq.~(\ref{eq:PPD_b_general}) are
\[\propto
M_b^{n_0-n}\exp\left(-M_b^{n_0-n}\right)\ll 1\]
where we have used $M_b=\exp(2\pi/\Omega_F)\gg 1$ (this estimate is
written for $p^{(n_0)}_{\rm opt}(t_f)Q_f/D= 1$).

A simple qualitative argument shows that the same $n_0$ gives a
major contribution to the sum in the numerator of
Eq.~(\ref{eq:PPD_b_general}). Indeed, the terms with different $n$
correspond to $t$ changing by an integer number of the modulation
periods $\tau_F$, whereas in its central part (section C in
Fig.~\ref{fig:scheme}) an optimal escape trajectory lasts for the
time small compared to $\tau_F$. Therefore
$|t-t_f|\ll \tau_F$. The formal condition for this approximation to
be true is that the Suzuki time $t_S$ (\ref{eq:Suzuki_time}) is
small compared to the modulation period.

The maximum of the Gaussian PPD peak (\ref{eq:PPD_b_large_theta_ad})
lies at $Q_{n_0}(t|q_f,t_f)$. This function is a sum of the optimal
path $Q^{(n)}_{\rm opt}$, which is located inside the attraction
basin, and the decaying in time term $\propto Q_f-Q^{(n)}_{\rm
opt}(t_f)$, which is determined by the final point $(q_f,t_f)$ and
is located outside the attraction basin. Its time dependence is
particularly simple in the case where $t_f$ corresponds to the
maximum of the distribution $\rho(q_f,t_f)$, i.e., $p_{\rm
opt}^{(n_0)}(t_f)=D/Q_f$,
\begin{equation}
\label{eq:maximum_Gauss_b}
Q_{n_0}(t|Q_f,t_f)\approx Q_fe^{-\mu_{b{\rm m}}(t_f-t)}
-\frac{D}{\mu_{b{\rm m}}Q_f}e^{\mu_{b{\rm m}}(t_f-t)}.
\end{equation}
\begin{center}
\begin{figure}[h]
\includegraphics[width=3.4in]{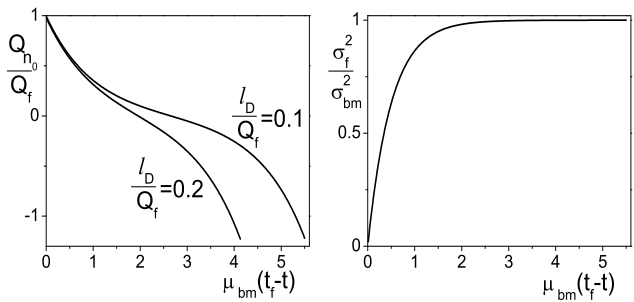}
\caption{(Color online) The reduced position of the maximum (left)
and the reduced width (right) of the PPD
(\ref{eq:PPD_b_large_theta_ad}) in the adiabatic regime as a
function of the reduced time $\mu_{b{\rm m}}(t_f-t)$.}
\label{fig:Gauss_parameters}
\end{figure}
\end{center}
The first term in $Q_{n_0}$, Eq.~(\ref{eq:maximum_Gauss_b}),
decreases with increasing $t_f-t$, which describes approaching the
basin boundary backward in time.  The second term, on the other
hand, increases with $t_f-t$; this term describes the motion,
backward in time, from the boundary to the interior of the
attraction basin. The motion at the boundary is initiated by noise,
and therefore this term is $\propto D$. For sufficiently long
$t_f-t$ the distribution maximum $Q_{n_0}$ approaches the optimal
path $Q^{(n_0)}_{\rm opt}(t)$. The overall behavior of $Q_{n_0}$ is
shown in Fig.~\ref{fig:Gauss_parameters}. It is seen that the motion
is slowed down near the basin boundary, and the slowing down
strongly depends on the noise intensity, in agreement with
Eq.~(\ref{eq:maximum_Gauss_b}).

The width of the Gaussian PPD peak (\ref{eq:PPD_b_large_theta_ad})
is diffusion-limited; its time dependence is given by
Eq.~(\ref{eq:ad_functions}) and is also simple. It is shown in the
right panel of Fig.~\ref{fig:Gauss_parameters}.

The condition that the integrands over $p$ in
Eq.~(\ref{eq:PPD_b_general}) are maximal for $p_{\rm opt}^{(n_0)}$
imposes a limitation on $Q$ where the explicit expression for the
PPD (\ref{eq:PPD_b_large_theta_ad}) applies. For $Q$ close to the
maximum of the distribution, $|Q-Q_{n_0}| \lesssim l_D$, this
condition takes the form $\theta \gg |Q_{\rm opt}^{(n_0)}(t)/l_D|$.
Therefore expression (\ref{eq:PPD_b_large_theta_ad}) describes the
distribution not only in the whole range between $(Q_f,t_f)$ and the
basin boundary, but also throughout the diffusion region around the
basin boundary and a region that goes deeper into the attraction
basin.

\section{Nonadiabatic regime near the basin boundary}
\label{sec:nonadiabatic}

The PPD can display a qualitatively different behavior for
nonadiabatic modulation, $\Omega_F \sim 1$. With increasing $t_f-t$
the PPD inside the attraction basin can split into several peaks.
Such splitting occurs already close to the basin boundary and is
described by the general expression (\ref{eq:PPD_b_general}).
Formally, the PPD has well-resolved multiple peaks when several
terms in the numerator of Eq.~(\ref{eq:PPD_b_general}) are of the
same order of magnitude.

In the nonadiabatic regime necessarily $\theta\gg 1$. Therefore
integration over $p$ in Eq.~(\ref{eq:PPD_b_general}) can be done by
the steepest descent method. The integrands are maximal for $p=
p^{(n)}_{\rm opt}$. Using Eq.~(\ref{eq:sigma_f_and_sigma_b}) one can
show that each term in the numerator gives a Gaussian distribution
over $Q$ centered at $Q_n\equiv Q_n(t|q_f,t_f)$,
Eq.~(\ref{eq:Q_n_0}), with variance $D\sigma_f^2(t_f,t)$,
Eq.~(\ref{eq:Sigmab_def}). This is similar to the adiabatic case
(\ref{eq:PPD_b_large_theta_ad}), except that now the PPD is a sum of
appropriately weighted Gaussian peaks,
\begin{eqnarray}
\label{eq:PPD_b_nonad} p_h(q,t|q_f,t_f)
=\frac{\sum_nB_nG\left(Q-Q_n; \sigma_f(t_f,t)\right)}{\sum_nB_n}.
\end{eqnarray}
The weighting factors $B_n$ are
\begin{eqnarray}
\label{eq:G_n_b} B_n\equiv B_n(q_f,t_f) = x_n\exp(-x_n),\nonumber\\
x_n=x_n(q_f,t_f)=p^{(n)}_{\rm opt}(t_f)Q_f/D.
\end{eqnarray}

If we keep only one term in the numerator and denominator in
Eq.~(\ref{eq:PPD_b_nonad}), with the same $n=n_0$ given by the
condition that $B_n$ is maximal for $n=n_0$,
Eq.~(\ref{eq:PPD_b_nonad}) goes over into
Eq.~(\ref{eq:PPD_b_large_theta_ad}). However, in the nonadiabatic
case there are regimes where the PPD as a function of $(Q,t)$ may
display several peaks. Eq.~(\ref{eq:PPD_b_nonad}) gives their shapes
near the maxima.

The number of peaks of $p_h$ depends on $Q,t$, and $\Omega_F$, as
well as the final point $(q_f,t_f)$. As before, we choose the final
point so that the probability density behind the basin boundary
$\rho(q_f,t_f)$ is close to its maximum over $t_f$, which occurs for
$x_{n_0}(q_f,t_f)\sim 1$. The amplitudes of different peaks are
\begin{equation}
\label{eq:peak_amplitudes}
B_{n_0+k}=x_{n_0}M_b^{-k}\exp\left(-x_{n_0}M_b^{-k}\right),
\end{equation}
with $M_b=\exp(2\pi/\Omega_F)$ being the Floquet multiplier,
Eq.~(\ref{eq:Floquet}). In practice, in the whole range of
modulation parameters where escape is strongly synchronized the
factor $M_b$ is comparatively large. Therefore for $x_{n_0}\sim 1$
the coefficients $B_{n_0+k}$ rapidly decay with increasing $|k|$,
and only few peaks of the PPD can be simultaneously observed,
primarily with non-negative $k=0,1,\ldots$.

The other condition for observing several peaks is that the distance
between them exceed their width, that is in particular
$|Q_{n_0}-Q_{n_0+1}|\gg l_D$. For $Q\approx Q_{n_0}\gg l_D$, i.e.,
when the PPD maximum is well behind the basin boundary with respect
to the attraction basin, we have $|Q_{n_0}-Q_{n_0+1}|\sim
|Q^{(n_0)}_{\rm opt}(t)| \sim l_D^2\kappa_b(t_f,t)/Q_f\ll l_D$,
where we have used that $Q_f\gg l_D$ and that $\kappa_b(t_f,t)\sim
1$ when the PPD maximum is behind the boundary. Therefore in this
region the PPD has indeed only one peak.

For longer $t_f-t$ when $\kappa_b(t_f,t)\gg 1$ and the peak
positions $Q_{n_0}, Q_{n_0+1}$ are well inside the basin of
attraction, the distance between them exceeds $l_D$. In this range
the corresponding peaks of $p_h$ can be resolved. The ratio of
their amplitudes is given by the factor $M_b$, as seen from
Eq.~(\ref{eq:peak_amplitudes}). This is in good agreement with the
results of numerical simulations for a specific model shown in
Fig.~\ref{fig:3D}, as will be discussed below.

We note that for $\Omega_F\gg 1$ synchronization of escape by
modulation becomes weak. Although the system is far from the
adiabatic limit, dynamics of the period-average coordinate is
similar to dynamics in the case of a stationary system. Even though
the factor $M_b$ becomes of order 1, of interest is the PPD with
respect to the period-averaged coordinate, and this PPD does not
display multiple peaks.

\section{The PPD inside the attraction basin}
\label{sec:PPD_inside}

We assume throughout this Section that escape is strongly
synchronized and $\theta\gg 1$. To find the PPD for $(q,t)$ deep
inside the attraction basin we will use the equation
\begin{equation}
\label{eq:PPD_match}
p_h(q,t|q_f,t_f)=\int dq' p_h(q,t|q',t')p_h(q',t'|q_f,t_f).
\end{equation}
It applies for $t < t' < t_f$, as follows from the definition
(\ref{eq:PPD_def}), and can be obtained \cite{Hales2000} from the
Chapman-Kolmogorov equation for the transition probability
$\rho(q_f,t_f|q,t)$.

As we show, $t'$ in Eq.~(\ref{eq:PPD_match}) can be chosen in such a
way that only a narrow range of $q'$ contributes to the integral.
The corresponding $q'$ are close but not too close to $q_b(t')$. In
this range both factors in the integrand, and hence the integral as
a whole, can be explicitly calculated. In particular, the function
$p_h(q',t'|q_f,t_f)$ is given by Eq.~(\ref{eq:PPD_b_nonad}).

The function $p_h(q,t|q',t')$ is the PPD inside the attraction
basin. For a periodically modulated system it was found earlier
\cite{Dykman1996a}. Because in escape the system is likely to move
close to one of the periodically repeated most probable escape paths
$q_{\rm opt}^{(n)}(t)$, we are interested in $p_h(q,t|q',t')$ for
both $(q,t)$ and $(q',t')$ lying close to such a path.  If both
$(q,t)$ and $(q',t')$ are close to an $n$th path $q_{\rm
opt}^{(n)}(t)$, the corresponding PPD $p_h^{(n)}(q,t|q',t')$ is
Gaussian \cite{Dykman1996a},
\begin{eqnarray}
\label{eq:PPD_inside} p_h^{(n)}(q,t|q',t')=
G\bigl(q-q_n(t|q',t');\sigma_n(t',t)\bigr)
\end{eqnarray}
Here $q_n(t|q',t')$ is the value of the coordinate at time $t$ on
the optimal path that leads to the point $(q',t')$. This optimal
path is described by Eq.~(\ref{eq:aux_system}).

The coordinate $q_n(t|q',t')$ as a function of time is close to
$q^{(n)}_{\rm opt}(t)$. We can seek it in the form
$q_{n}(t|q',t')=q_{\rm opt}^{(n)}(t)+\delta q_n(t|q',t')$. To the
lowest order in the deviation of $(q',t')$ from $q^{(n)}_{\rm
opt}(t')$, the function $\delta q_n$ can be found from linearized
equations (\ref{eq:aux_system}). This gives
\begin{eqnarray}
\label{eq:opt_linear}
&&\delta\ddot{q}_n = V_n(t)\delta q_n,\\
&&V_n(t)=\left[\partial_{qt}K+\frac{1}{2}\partial_{qq}(K^2)
\right]_{q_{\rm opt}^{(n)}(t)}.\nonumber
\end{eqnarray}
The boundary conditions for $\delta q_n$ are $\delta q_n(t'|q',t')
= q'-q_{\rm opt}^{(n)}(t')$ and $\lim\nolimits_{t\to-\infty}\delta
q_n(t|q',t')=0$; the latter condition simply means that the
optimal path approaches the attractor $q_a(t)$ as $t\to -\infty$.

It follows from Eq.~(\ref{eq:opt_linear}) that function
$\beta_n(t)=\delta\dot{q}_n/\delta q_n$ satisfies a first-order
(Riccati) equation,
\begin{equation}
\label{eq:eq_beta}
\dot{\beta}_n+\beta_n^2 = V_n(t).
\end{equation}
For $q',q_{\rm opt}^{(n)}(t')$ both close to the basin boundary
$q_b(t')$, with account taken of the boundary conditions for
$\delta q_n$ the solution of Eq.~(\ref{eq:opt_linear}) can be
written in terms of $\beta_n$ as
\begin{equation}
\label{eq:deltaq} \delta q_n(t|q',t') = \left(Q'-Q_{\rm
opt}^{(n)}(t')\right)\kappa_n(t,t'),
\end{equation}
where
\begin{equation}
\label{eq:kappa_n} \kappa_n(t,t')=
\exp\!\!\left[\int_{t'}^{t}\!\!d\tau\beta_n(\tau)\right].
\end{equation}

We now discuss the asymptotic behavior of $\beta_n(t)$. For not too
large $t'-t$ the function $V_n(t)$ in Eqs.~(\ref{eq:opt_linear}),
(\ref{eq:eq_beta}) should be calculated for $q^{(n)}_{\rm
opt}(t)\approx q_b(t)$. For such $V_n(t)$, the solution $\delta
q_n(t)$ of Eq.~(\ref{eq:opt_linear}) can either exponentially
increase or decrease with increasing $t'-t$. Of interest to us is
the decreasing solution, which will ultimately go to zero for
$t'-t\to\infty$.  A simple calculation shows that for this solution
\begin{eqnarray*}
\beta_n(t)\approx \mu_b(t)\equiv [\partial_qK]_{q_b(t)}, \qquad
q_{\rm opt}^{(n)}(t)\approx q_b(t).
\end{eqnarray*}

For much larger $t'-t$ the optimal path $q_{\rm opt}^{(n)}(t)$
approaches the attractor $q_a(t)$, and then the function $V_n(t)$ in
Eqs.~(\ref{eq:opt_linear}), (\ref{eq:eq_beta}) should be calculated
for $q^{(n)}_{\rm opt}(t)\approx q_a(t)$. One can show that the
solution $\delta q_n(t)\to 0$ for $t\to -\infty$ corresponds to
\begin{eqnarray*}
\beta_n(t)\approx \mu_a(t)+ 2\sigma_a^{-2}(t), \qquad q_{\rm
opt}^{(n)}(t)\approx q_a(t),
\end{eqnarray*}
where $\sigma_a^2(t)$ is the variance of the periodic distribution
about the attractor given by Eq.~(\ref{eq:sigma_a}).

To the lowest order in $q'-q_{\rm opt}^{(n)}(t')$, the variance of
the prehistory probability distribution (\ref{eq:PPD_inside}) is
also expressed in terms of the function $\beta_n$
\cite{Dykman1996a},
\begin{equation}
\label{eq:var1}
\sigma^2_n(t',t) = 2\int_t^{t'}d\tau\kappa_n^{-2}(\tau,t).
\end{equation}
Using Eqs.~(\ref{eq:eq_beta}), (\ref{eq:kappa_n}), and
(\ref{eq:var1}) one can show that $\sigma_n^2(t',t\to
-\infty)=\sigma_a^2(t)$. For $t\to -\infty$ the distribution
$p_h^{(n)}(q,t|q',t')$ goes over into the periodic Gaussian
distribution $\rho_a(q,t)$, Eq.~(\ref{eq:PPD_limit}), centered at
the periodic attractor.

It follows from Eqs.~(\ref{eq:PPD_inside})-(\ref{eq:deltaq}) that,
for $q$ close to $q_{\rm opt}^{(n)}(t)$, the PPD
$p_h^{(n)}(q,t|q',t')$ displays a diffusion-broadened Gaussian peak
as a function of $q'$ with maximum close to $q_{\rm opt}^{(n)}(t')$.
The displacement of the maximum over $q'$ from $q_{\rm
opt}^{(n)}(t')$ is $\propto [q-q_{\rm opt}^{(n)}(t)]$. On the other
hand, the function $p_h(q',t'|q_f,t_f)$ as given by
Eq.~(\ref{eq:PPD_b_nonad}) is a sum of Gaussian distributions
centered at the optimal escape paths. Therefore integration over
$q'$ in Eq.~(\ref{eq:PPD_match}) can be done by the steepest descent
method. The extreme values of $q'$ are also close to the optimal
escape paths.

Different peaks of the PPD $p_h(q',t'|q_f,t_f)$,
Eq.~(\ref{eq:PPD_b_nonad}), correspond to the MPEP's shifted by an
integer number of modulation periods. For each of these peaks one
should use in Eq.~(\ref{eq:PPD_match}) the PPD $p_h(q,t|q't')$ given
by $p_h^{(n)}$ with the appropriate $n$.

The result of integration over $q'$ describes the peaks of the PPD
$p_h(q,t|q_f,t_f)$ for $q$ close to $q_{\rm opt}^{(n)}(t)$ with
appropriate $n$. These peaks are Gaussian near the maxima,
\begin{widetext}
\begin{equation}
\label{eq:PPD_final}
p_h(q,t|q_f,t_f)=\frac{1}{\sum_nB_n(q_f,t_f)}\sum_nB_n(q_f,t_f)G\bigl(q-q_{\rm
opt}^{(n)}(t)-\left(Q_f-Q_{\rm
opt}^{(n)}(t_f)\right)\kappa_n^{-1}(t_f,t);\sigma_n(t_f,t)\bigr).
\end{equation}
\end{widetext}
In obtaining this equation we used the relations
\begin{eqnarray*}
&&\kappa_b(t_f,t')\kappa_n(t',t)=\kappa_n(t_f,t),\\
&&\sigma_n^2(t',t)+\sigma_f^2(t_f,t')\kappa_n^{-2}(t',t)=\sigma_n^2(t_f,t),
\end{eqnarray*}
which in turn follow from the relation $\beta_n(t)\approx \mu_b(t)$
in the harmonic region near the basin boundary. As a consequence,
the result of integration over $q'$ in Eq.~(\ref{eq:PPD_match}) is
independent of the matching time $t'$.

Equation (\ref{eq:PPD_final}) is the central result of the present
paper. It gives, in the explicit form, the PPD of finding the system
at a position $q$ inside the attraction basin at a moment $t$,
provided the system has been observed at a point $q_f$ outside the
attraction basin at a time $t_f>t$. It shows that, in the regime of
strong synchronization, the peaks are Gaussian. They are centered at
the most probable escape paths and are diffusion broadened
throughout the attraction basin. This is in dramatic contrast with
the PPD in the absence of synchronization, where the PPD peak inside
the attraction basin is strongly asymmetric and its width is
independent of the noise intensity $D$ \cite{Hales2000}.

It follows from Eq.~(\ref{eq:PPD_final}) that the PPD may have
multiple peaks inside the attraction basin. They can be observed
only for strongly nonadiabatic modulation, where the Suzuki time is
$t_S\gtrsim \tau_F$ . In this case the system stays in the diffusion
layer around the basin boundary $q_b(t)$ long enough to accumulate
influxes from several periodically repeated most probable escape
paths $q^{(n)}_{\rm opt}(t)$. On the other hand, the modulation
period $\tau_F$ should not be too short, because the strong
synchronization of escape would be lost. Since synchronization loss
occurs for $t_{\rm r}\gg \tau_F$, the PPD displays well resolved
multiple peaks only in a limited parameter range. The limitation is
more restrictive to the considered case where the final point
$(q_f,t_f)$ is close to the maximum of the distribution behind the
barrier. If this condition is not imposed, the peaks are well
resolved in a broader range, but measuring the PPD becomes more
complicated on the whole. An example is discussed in the next
Section.

An important feature of the distribution (\ref{eq:PPD_final}) is
weak dependence of the shape of the Gaussian peaks inside the
attraction basin on the final point $(q_f,t_f)$, which is a
consequence of the smallness of the factor $\kappa^{-1}(t_f,t)$.
This shows that the PPD reveals the actual structure of the tubes of
the paths followed in escape inside the attraction basin. In
contrast, the relative amplitudes of the PPD peaks $B_n$ are
sensitive to the choice of the point $(q_f,t_f)$.

\section{Results for a model system}
\label{sec:model_system}

In this section, we present the results of numerical simulations of
the PPD for a simple model system and compare them with the
analytical predictions. We consider a Brownian particle in a
sinusoidally modulated potential of the form of a cubic parabola.
The Langevin equation of motion has the form of
Eq.~(\ref{eq:Langevin}) with
\begin{equation}
\label{eq:eom_model} K(q,t) = q^2 - \frac{1}{4} +
A\cos(\omega_Ft).
\end{equation}

The dynamics was simulated using the second-order integration scheme
for stochastic differential equations \cite{Mannella2002}. The
system was initially prepared in the vicinity of the metastable
state $q_a(t)$. The final point $(q_f,t_f)$ was chosen behind the
basin boundary $q_b(t)$. The PPD was calculated as a normalized
probability distribution of paths $q(t)$ arriving at the point $q_f$
for a particular modulation phase $\phi_f=\omega_Ft_f$(mod\,$2\pi$).

An example of the full PPD is shown in Fig.~\ref{fig:3D}. The point
$(q_f,t_f)$ is chosen so that in escape the system is likely to pass
near it ($t_f$ is determined ${\rm mod}\, \tau_F$), that is the
quasistationary distribution $\rho(q_f,t_f)$ is close to its maximum
over $t_f$ for a given $q_f$ behind the basin boundary [the
parameter $p^{(n_0)}_{\rm opt}(t_f)[q_f-q_b(t_f)]/D$ is equal to
$1.2$, whereas the maximum of $\rho(q_f,t_f)$ is expected where this
parameter is equal to 1, \cite{Ryvkine2005}]. For the chosen
modulation parameters and noise intensity the calculated activation
energy of escape is $R\approx 0.0910$ and $R/D\approx 9.1$; the
ratio of the modulation frequency to the relaxation rate
$\Omega_F\approx 2.24$. We accumulated $\sim 10^5$ escape
trajectories that arrive into a small area centered at $(q_f,t_f)$,
with width $\delta q=0.02, \delta t=0.02\tau_F$.

It is seen from Fig.~\ref{fig:3D} that, in the regime of strong
synchronization and for strongly non-Gaussian pulses of escape
current, the peaks of the PPD are narrow both behind the basin
boundary and inside the attraction basin. Moreover, for the chosen
parameter values two distinct peaks of the PPD are well resolved
inside the attraction basin. They correspond to the two paths the
system is most likely to follow in escape.

\begin{figure}[ht]
\includegraphics[width=3.0in]{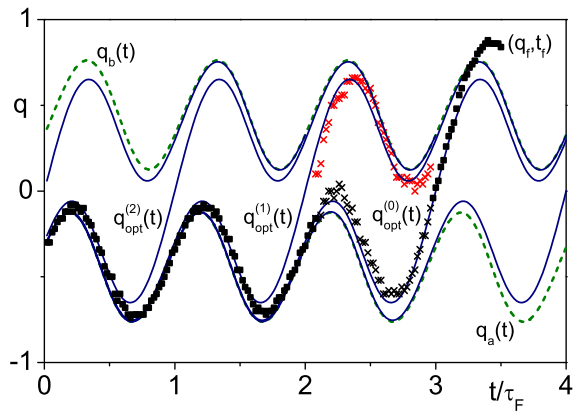}
\caption{(Color online) The positions of the maxima of the PPD
$p_h(q,t|q_f,t_f)$ in Fig.~\ref{fig:3D}, which show the most
probable paths followed by the system in escape. The data of
simulations are shown by full squares where the PPD has one peak and
by crosses where two peaks are well resolved. Solid lines show
periodically repeated most probable escape paths $q^{(n)}_{\rm
opt}(t)$ for the model (\ref{eq:Langevin}), (\ref{eq:eom_model}).
Dashed lines show the basin boundary $q_b(t)$ and the attractor
$q_a(t)$. } \label{fig:maxima}
\end{figure}

The positions of the PPD peaks on $(q,t)$ plane are shown in
Fig.~\ref{fig:maxima} with full squares where there is one peak, and
with crosses where two peaks are well resolved. They are compared
with the periodically repeated optimal escape paths calculated by
solving numerically the variational equations (\ref{eq:aux_system})
for the model (\ref{eq:Langevin}), (\ref{eq:eom_model}). Such paths
start from the periodic attractor $q_a(t)$ for $t\to -\infty$ and
approach the basin boundary $q_b(t)$ for $t\to\infty$, which are
also shown in the figure. It is seen that, as expected from our
analysis, the PPD maxima lie nearly on top of the most probable
escape paths. The small deviation is due to diffusion broadening and
associated small asymmetry of the PPD peaks for the noise intensity
used in the simulations. Fig.~\ref{fig:maxima} demonstrates that
studying the PPD provides a direct way of observing most probable
escape paths.

\begin{figure}[ht]
\includegraphics[width=3.2in]{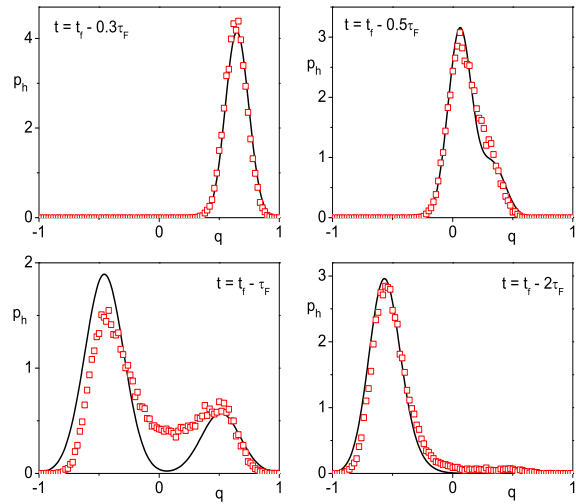}
\caption{(Color online) Cross-sections of the PPD $p_h(q,t|q_f,t_f)$
shown in Fig.~\ref{fig:3D} as functions of the coordinate $q$ for
the time $t_f-t= 0.3\tau_F, 0.5\tau_F, \tau_F, 2\tau_F$. The results
refer to the model system (\ref{eq:Langevin}), (\ref{eq:eom_model})
with $A=0.7$, $\omega_F=2$, $D=0.01$, $q_f=0.8$, $t_f = 0.5\tau_F$.
The point $(q_f,t_f)$ is close to the expected maximum of the
distribution behind the basin boundary. Solid lines show the
expression (\ref{eq:PPD_final}). Squares show the results of
simulations.
 }
\label{fig:profiles}
\end{figure}

The observed shape of the PPD peaks is compared with the theory in
Fig.~\ref{fig:profiles},  which shows the cross-sections of the PPD
at several instants of time counted off from the final time $t_f$.
For small $t_f-t$ the system is behind the basin boundary and moves
close to the noise-free trajectory leading to $(q_f,t_f)$ . The top
left panel refers to the case where the system is localized close to
the boundary. Here and for smaller $t_f-t$ the PPD has a single
sharp peak.

For earlier time (larger $t_f-t$) the system could either be moving
towards the basin boundary along the most probable escape path or
could have been fluctuating about the basin boundary after it had
arrived to its vicinity along the previous MPEP (shifted by
$\tau_F$). As we showed analytically, the probability of staying
near the boundary is smaller, but the PPD may still display two
peaks. This is seen in the right top panel. The main peak
corresponds to motion along the MPEP, which is close to $q_b(t)$ for
the chosen time. The higher-$q$ shoulder corresponds to the poorly
resolved (for the chosen time) peak for fluctuations about the basin
boundary.

For still larger but not too large $t_f-t$ the escaping system
should have been moving towards the basin boundary. In the present
case it most likely followed one of the two periodically repeated
MPEPs, with different probabilities. Well-resolved peaks of the PPD
in this range are seen in the left lower panel of
Fig.~\ref{fig:profiles}. For large $t_f-t$ compared to the
relaxation time and $\tau_F$, the system should have been
fluctuating about the attractor $q_a(t)$. The PPD in this case
should have a single peak, which is seen in the right lower panel.

It is seen from Fig.~\ref{fig:profiles} that the results of
simulations agree with the analytical results. Not surprisingly, the
observed peaks are broader than the asymptotic theory predicts. This
is a consequence of the relatively large noise intensity used in the
simulations.

\section{Conclusions}
\label{sec:conclusions}

In this paper we have studied the prehistory probability
distribution for activated escape in periodically modulated systems.
We have shown that the PPD as a function of coordinates can display
one or several narrow peaks within the basin of attraction to the
metastable state. They correspond to narrow ridges of the PPD in
$(q,t)$ space. The ridges are centered at the periodically repeated
most probable escape paths. Their cross-sections are Gaussian near
the maxima and are diffusion broadened, see
Eq.~(\ref{eq:PPD_final}).

The number of the PPD peaks that can be observed depends on the
parameters of the system. The multi-peak structure is best resolved
in a limited parameter range. On the one hand, the modulation should
not be too slow, so that the system does not follow it
adiabatically. On the other hand, it should not be too fast, so that
escape events are strongly synchronized and the system dynamics is
not described by period-averaged coordinates.

Most of the results of the paper refer to the case where the final
point $(q_f,t_f)$ is chosen so that the escaped system has a high
probability density of passing through this point. Such choice
simplifies the experimental observation of the PPD. For the
corresponding $(q_f,t_f)$ the amplitudes of different peaks of the
PPD [heights of the ridges in $(q,t)$ space] differ from each other
significantly, as seen from Eq.~(\ref{eq:peak_amplitudes}) and
Figs.~\ref{fig:3D}, \ref{fig:profiles}. These amplitudes are
sensitive to the choice of $(q_f,t_f)$. In contrast, the positions
and shapes of the PPD peaks very weakly depend on $(q_f,t_f)$. This
shows that the PPD provides a means for studying the distribution of
trajectories leading to escape.

We have performed extensive numerical simulations of escape for a
model Brownian particle in a modulated potential well. The
simulations confirm the possibility to clearly observe most probable
escape paths. They are in a good qualitative and quantitative
agreement with the analytical theory.

Observing most probable escape paths is not only interesting on its
own, but also has broader implications. First, periodically
modulated systems are an important class of systems far from thermal
equilibrium. In contrast to the case of equilibrium systems, optimal
fluctuational paths in nonequilibrium systems have no immediate
relation to dynamical trajectories in the absence of noise; in
particular, they may not be obtained by just reversing time. The
MPEP's can display interesting and counterintuitive behavior
\cite{Maier1993a,Dykman1990}, and studying them provides an insight
into general features of dynamics away from thermal equilibrium.
Second, understanding the dynamics of a system in escape paves the
way for efficient control of this dynamics and the escape
probability itself; as will be discussed in a separate publication,
such control can be accomplished by comparatively weak field pulses
applied in the right place and at the right time.

In conclusion, the results of this paper suggest a way of direct
observation of most probable escape paths, in space and time. They
also describe, qualitatively and quantitatively, the distribution of
the trajectories followed in escape.

\section*{ACKNOWLEDGMENTS}
This research has been supported in part by the NSF Grant No.
DMR-0305746.

\end{document}